\documentclass[preprint,nofootinbib,aps,superscriptaddress,floatfix]{revtex4}
\usepackage{amsmath}
\usepackage[dvips]{graphics,color}
\begin{document}

\def\LQCD{\Lambda_{\rm QCD}}
\def\lqcd{\Lambda_{\rm QCD}}
\def\xslash#1{{\rlap{$#1$}/}}
\newcommand{\orderalpha}{ {\cal{O}}(\alpha_s)}
\newcommand{\orderalphasqr}{ {\cal{O}}(\alpha_s^2)}
\newcommand{\btoc}{\bar{B} \to \rm X_c \, \ell \, \bar{\nu}}
\newcommand{\nn}{\nonumber}
\newcommand\beq{\begin{equation}}
\newcommand\eeq{\end{equation}}
\newcommand\Tr{\text{Tr}}
\newcommand\vev[1]{\langle #1\rangle}
\newcommand\diag{\text{diag}}
\newcommand\hc{\text{h.c.}}
\newcommand\LLH{$\text{L}^2\text{H}$}
\def\be{\begin{equation}}
\def\ee{\end{equation}}
\def\ba{\begin{eqnarray}}
\def\ea{\end{eqnarray}}
\def\ge{\mathrel{\raise.3ex\hbox{$>$\kern-.75em\lower1ex\hbox{$\sim$}}}}
\def\la{\mathrel{\raise.3ex\hbox{$<$\kern-.75em\lower1ex\hbox{$\sim$}}}}
\newcommand{\sect}[1]{\section{#1}\setcounter{equation}{0}}
\def\thesection{\arabic{section}}
\def\theequation{\arabic{equation}}
\def\simgt{\mathrel{\raise.3ex\hbox{$>$\kern-.75em\lower1ex\hbox{$\sim$}}}}
\def\simlt{\mathrel{\raise.3ex\hbox{$<$\kern-.75em\lower1ex\hbox{$\sim$}}}}
\newcommand{\s}{\mbox{$\sigma$}}
\newcommand{\bi}[1]{\bibitem{#1}}
\newcommand{\fr}[2]{\frac{#1}{#2}}
\newcommand{\gm}{\mbox{$\gamma_{\mu}$}}
\newcommand{\gn}{\mbox{$\gamma_{\nu}$}}
\newcommand{\Le}{\mbox{$\fr{1+\gamma_5}{2}$}}
\newcommand{\R}{\mbox{$\fr{1-\gamma_5}{2}$}}
\newcommand{\GD}{\mbox{$\tilde{G}$}}
\newcommand{\gf}{\mbox{$\gamma_{5}$}}
\newcommand{\tb}{\tan\beta}
\newcommand{\Ima}{\mbox{Im}}
\newcommand{\Rea}{\mbox{Re}}
\newcommand{\psl}{\slash{\!\!\!p}}
\newcommand{\dsl}{\slash{\partial}}
\newcommand{\cp}{\;\;\slash{\!\!\!\!\!\!\rm CP}}
\newcommand{\qq}{\langle \ov{q}q\rangle}
\newcommand{\uGu}{\bar{u}g_s(G\si) u}
\newcommand{\dGd}{\bar{d}g_s(G\si) d}
\newcommand{\nc}{\newcommand}
\newcommand{\uu}{\bar{u}u}
\newcommand{\dd}{\bar{d}d}
\nc{\gone}{\bar g_{\pi NN}^{(1)}}
\nc{\gzero}{\bar g_{\pi NN}^{(0)}}
\nc{\al}{\alpha}
\nc{\ga}{\gamma}
\nc{\de}{\delta}
\nc{\ep}{\epsilon}
\nc{\ze}{\zeta}
\nc{\et}{\eta}
\nc{\ka}{\kappa}
\nc{\rh}{\rho}
\nc{\si}{\sigma}
\nc{\ta}{\tau}
\nc{\up}{\upsilon}
\nc{\ph}{\phi}
\nc{\ch}{\chi}
\nc{\ps}{\psi}
\nc{\om}{\omega}
\nc{\Ga}{\Gamma}
\nc{\De}{\Delta}
\nc{\La}{\Lambda}
\nc{\Si}{\Sigma}
\nc{\Up}{\Upsilon}
\nc{\Ph}{\Phi}
\nc{\Ps}{\Psi}
\nc{\Om}{\Omega}
\nc{\ptl}{\partial}
\nc{\del}{\nabla}
\nc{\ov}{\overline}
\nc{\newcaption}[1]{\centerline{\parbox{15cm}{\caption{#1}}}}

\def\beq{\begin{equation}}
\def\eeq{\end{equation}}
\def\bmat{\begin{displaymath}}
\def\emat{\end{displaymath}}
\def\bear{\begin{eqnarray}}
\def\eear{\end{eqnarray}}
\def\ba{\begin{eqnarray}}
\def\ea{\end{eqnarray}}
\def\bery{\begin{array}}
\def\ery{\end{array}}
\def\bit{\begin{itemize}}
\def\eit{\end{itemize}}
\def\ben{\begin{enumerate}}
\def\een{\end{enumerate}}
\def\btab{\begin{tabular}}
\def\etab{\end{tabular}}
\def\btbl{\begin{table}}
\def\etbl{\end{table}}
\def\bfig{\begin{figure}[htb]}
\def\efig{\end{figure}}
\def\bpic{\begin{picture}}
\def\epic{\end{picture}}

\def\st{\scriptstyle}
\def\ss{\scriptscriptstyle}
\def\hsx{\hspace{0.06in}}
\def\hse{\hspace{0.08in}}
\def\hst{\hspace{0.12in}}
\def\nnl{\nonumber \\}
\def\nl{\nonumber \\ &&}

\def\hocm{\hspace{1cm}}
\def\htcm{\hspace{2cm}}

\def\ga{\mathrel{\raise.3ex\hbox{$>$\kern-.75em\lower1ex\hbox{$\sim$}}}}
\def\la{\mathrel{\raise.3ex\hbox{$<$\kern-.75em\lower1ex\hbox{$\sim$}}}}
\def\gappeq{\mathrel{\rlap {\raise.5ex\hbox{$>$}}
{\lower.5ex\hbox{$\sim$}}}}
\def\lappeq{\mathrel{\rlap{\raise.5ex\hbox{$<$}}
{\lower.5ex\hbox{$\sim$}}}}
\def\ohsq{\Omega_{\widetilde\chi}\, h^2}
\def\gyr{{\rm \, G\kern-0.125em yr}}
\def\mev{{\rm \, Me\kern-0.125em V}}
\def\gev{{\rm \, Ge\kern-0.125em V}}
\def\tev{{\rm \, Te\kern-0.125em V}}
\def\cp{C\!P}
\def\tsq{|{\cal T}|^2}
\def\halft{{\textstyle{1\over2}}}
\def\slash#1{\rlap{\hbox{$\mskip 1 mu /$}}#1}%
\def\tbt{\tan \beta}
\def\ttbt{\tan^2 \beta}
\def\hc{{\rm h.c.}}
\def\emunu{\eta^{\hspace{0.01in} \mu \hspace{0.01in} \nu}}
\def\bfp{{\bf p}}
\def\nhat{{\bf \hat{n}}}

\def\half{\frac{1}{2}}
\def\athird{\frac{1}{3}}
\def\aforth{\frac{1}{4}}
\def\Tr{\rm Tr}
\def\Ker{\rm Ker}
\def\index{\rm index}
\def\bmtheta{\mbox{\boldmath $\theta$}}
\def\bmphi{\mbox{\boldmath $\phi$}}
\def\bmalpha{\mbox{\boldmath $\alpha$}}
\def\bmsigma{\mbox{\boldmath $\sigma$}}
\def\bmgamma{\mbox{\boldmath $\gamma$}}
\def\bmomega{\mbox{\boldmath $\omega$}}

\newcommand{\FRAME}[1]{\fbox{\mbox{$#1$}}}

\title{$R$-parity preserving super-WIMP decays}

\author{Maxim Pospelov$^{1,2}$ and 
Michael Trott}\thanks{pospelov@perimeterinstitute.ca \\ mtrott@perimeterinstitute.ca}
\affiliation{{\it Perimeter Institute for Theoretical Physics, Waterloo,
Ontario N2J 2W9, Canada}
\\
$^2${\it Department of Physics and Astronomy, University of Victoria, 
     Victoria, BC, V8P 1A1 Canada}
}
\date{\today}            
\begin{abstract}
We point out that when the decay of one electroweak scale super-WIMP state to another 
occurs at second order in a super-weak coupling constant,
this can naturally lead to decay lifetimes that are much larger than the 
age of the Universe, and create observable consequences for the indirect detection 
of dark matter. We demonstrate this in a supersymmetric model with Dirac neutrinos, 
where the right-handed scalar neutrinos
are the lightest and next-to-lightest supersymmetric partners. 
We show that this model produces a super-WIMP decay rate scaling as $m_{\nu}^4/({\rm  weak~scale})^3$, 
and may significantly enhance the fraction of  energetic electrons and positrons 
over anti-protons in the decay products.
Such a signature is consistent with
the observations recently reported by the PAMELA experiment.
 

\end{abstract}

\maketitle
\section{Introduction}

The detection of non-gravitational cold dark matter (DM) interactions is an important goal that
drives some aspects of modern particle physics, cosmology and astrophysics. The well-measured abundance 
of dark matter \cite{Komatsu:2008hk} does not specify the origin or specific nature of DM. 
In particle physics, when the DM particle masses are comparable 
to the electroweak scale ($m_\chi \sim {\it v}$) there are two distinct (well-motivated) possibilities.\footnote{Of course, it is also possible that nature could correspond to 
the parameter space between two extremes of WIMPs and super-WIMPs and/or have multiple DM states of each type.} 
The first one, is the weakly interacting massive particle (WIMP)
framework, in which the initially thermalized abundance of WIMPs is 
reduced via their weak annihilation, starting at the particles freeze 
out temperatures $T_f\sim 0.05 \, m_\chi$ \cite{Vysotsky:1977pe,Lee:1977ua}. The second possibility 
(the super-WIMP) postulates that the DM particles have an interaction rate
{\it much} weaker than the weak interactions. In this case, one assumes that the initial cosmological abundance 
of such particles is small, 
and that the rate for their direct production from thermal SM states, 
controlled by the square of the super-WIMP coupling $y_{SW}$, remains 
small relative to the Hubble expansion rate throughout the entire history of the Universe. 
The ratio of the super-WIMP thermalization rate to the Hubble rate at the weak scale, 
$\Gamma_{th}/H\sim 10^{-4} \, y_{SW}^2\times (M_{Pl}/{\rm weak ~ scale})$,  
gives a crude estimate for the resulting number densities of elecroweak scale super-WIMPs 
weighted by entropy. Observations require this ratio be $\lesssim 10^{-12}$,
which sets the benchmark value for the super-WIMP coupling, $y_{SW} \la 10^{-12}$. The stability of
electroweak scale super-WIMPs would have to be ensured by some parity in the dark sector.

While the WIMP framework gives hope to the goal of direct detection of DM \cite{Jungman:1995df,Bergstrom:2000pn,Bertone:2004pz}  (see however \cite{Pospelov:2007mp}), 
electroweak scale super-WIMPs would necessarily 
have a very tiny interaction strength that generally cannot lead to direct detection. 
Conversely, indirect signatures of DM, such as energetic gamma-rays,
positrons and anti-protons, may be created by either WIMPs or super-WIMPs, 
provided that the latter decay with a lifetime longer than the age of the Universe. 

The recent claim by the PAMELA collaboration\footnote{As well as the more recent claim by the ATIC 
collaboration of an electron excess at higher energies \cite{Torii:2008xu} 
and other past hints of energetic electron/positron excess in the galaxy at high energies 
\cite{Collaboration:2008aa,Beatty:2004cy,Aguilar:2007yf,Dobler:2007wv,Strong:2005zx}.}  
\cite{Adriani:2008zr} of an excess in the
positron fraction above 10 GeV, 
is broadly in agreement with expectations for WIMP annihilation 
\cite{Turner:1989kg,Kamionkowski:1990ty,Baltz:2001ir,Kane:2002nm,Hooper:2004bq,Cirelli:2007xd,Cholis:2008vb}
or WIMP/super-WIMP decay \cite{Buchmuller:2007ui,Ibarra:2008qg} leading to an excess postron flux above 
the background of secondary postron production. 
Even though an astrophyiscal origin of this signal (unrelated to dark matter) is not 
ruled out, there have been numerous attempts to link the PAMELA positron excess to dark matter.
Various analyses of the positron flux 
\cite{Cholis:2008hb,Yin:2008bs,Ishiwata:2008cv,Ibarra:2008jk}
find that the PAMELA result can be fit to models of dark matter, provided that the annihilation
and/or decay rate to positrons satisfy the following criteria:
\begin{eqnarray}
\label{annih}
{\rm WIMP~~annihilation}&:&~~~\langle \sigma v \rangle_{e^+} \sim O(3 \times 10^{-24} \, {\rm cm}^3s^{-1}) 
\times \left(\fr{m_\chi}{500~{\rm GeV}}\right)^2 \\
{\rm WIMP /super\!-\!WIMP ~~decay}&:&~~~ \Gamma_{e^+} \sim O(10^{-51}~{\rm GeV})\times  
\fr{m_\chi}{1~{\rm TeV}},
\label{decay}
\end{eqnarray}
where $m_\chi$ is the mass of decaying/annihilating particles.
One should keep in mind that these estimates have uncertainties 
both due to their dependence on the modeling of the propagation of positrons
in the galaxy (see \cite{Maurin:2001sj,Delahaye:2007fr} for a discussion) and due 
to uncertainties in the local dark matter energy density. 
In addition, PAMELA has also 
reported the measured flux of anti-protons \cite{Adriani:2008zq}, which is 
well described by standard astrophysical production of anti-protons
\cite{Donato:2008jk}. 

While the annihilation rate in Eqn. (\ref{annih}) of WIMPs inside our galaxy
is naively in conflict with the abundance-derived rate of 
$\langle \sigma v \rangle_{T_f} \simeq 3 \times 10^{-26} \, {\rm cm}^3s^{-1}$, 
it has been argued that this may not necessarily be the case
\cite{Cirelli:2008pk,ArkaniHamed:2008qn,Pospelov:2008jd}, 
as significant enhancement factors are possible due to 
resonance annihilation and/or $v^{-1}$ enhancement from the long-range attraction 
in the dark matter sector\footnote{Other recent works motivated by PAMELA data
include 
\cite{Finkbeiner:2008qu,Hooper:2008kg,Fairbairn:2008fb,Nelson:2008hj,Harnik:2008uu,Fox:2008kb,Ponton:2008zv,Pospelov:2008zw,Hamaguchi:2008rv,Zurek,Bai:2008jt}}. This can boost the galactic annihilation 
rate far above $\langle \sigma v \rangle_{T_f}$. 
As no excess in the anti-proton flux is reported, the
positron excess suggests mostly leptonic channels of annihilation 
and small rates of decay to quark-antiquark pairs for the DM. 

As a "boost factor" of size $10-10^3$ needed for the interpretation of the 
PAMELA results via WIMP annihilation may be accomplished in many ways 
through model-building, an explanation via
the decays of WIMPs or super-WIMPs may seem more {\em ad hoc}.  
In order to obtain
the decay width of $10^{-51}$ GeV  in Eqn.(\ref{decay}), one would typically introduce a 
decay constant $y_{decay} \sim 10^{-25}-10^{-23}$, such that $\Gamma_{decay} 
\propto y_{decay}^2\times({\rm weak~scale})/16 \pi^2$. Such a coupling is
twenty orders of magnitude smaller than a typical 
WIMP coupling and more than ten orders orders of magnitude smaller than the upper bound on the  super-WIMP 
coupling. Although possible, such an option is not appealing, yet
in most of the decay scenarios discussed in the literature, such a coupling is introduced by hand 
to fit the PAMELA data.

In this note, we consider the generic possibility when there are more than one super-WIMP
states present in the DM sector, and the decay of one state to another is kinematically allowed. 
Such a decay can happen only in the second order of the super-weak 
coupling constant, and can be {\em naturally} suppressed down to the level given by Eqn. (\ref{decay}):
\be
({\rm superWIMP})_1 \to  ({\rm superWIMP})_2 + {\rm SM~ particles}, ~~\Gamma \sim O(y_{SW}^4).
\ee 
The forth order of the superweak coupling, $O(y_{SW}^4)$, is necessary a tiny number not far 
from $10^{-50}$, which may lead to a natural realization of the $O(10^{-51}~{\rm GeV })$ decay width. 

We demonstrate this scenario realized in 
a supersymmetric Standard Model (MSSM) with exact $R$-parity conservation
extended with right-handed (RH) neutrino superfields and
Dirac (or nearly Dirac) neutrino masses.  In this model one 
identifies the Yukawa couplings in the neutrino sector $y_\nu\sim 10^{-(12 - 13)}$ with $y_{SW}$
\cite{McDonald:2006if,Asaka:2007zz,Asaka:2006fs}.
Analyzing the model, we find that the excess of energetic positrons and electrons
may naturally arize from the decay of one RH sneutrino species into another  
$\tilde{\nu}_R^1  \rightarrow \tilde{\nu}_R^2  \, \ell^+ \, \ell^-$, and that for certain 
domains of the parameter space the production of anti-protons is 
inhibited.  This particular realization of the decaying super-WIMP 
scenario is the subject of this paper.

\section{Naturally unstable SUSY DM candidates}

The spectrum of supersymmetric particles depends on 
the nature of supersymmetry breaking generating the soft masses 
of the MSSM Lagrangian.  
Currently, there is no overwhelming reason to adhere to any particular scheme of 
SUSY breaking; the only objective restriction on the 
SUSY-breaking mass pattern comes from the resulting phenomenology. Thus,
the charged lightest supersymmetric particle (LSP) is ruled out due to the desire 
for a natural DM candidate, and a nearly 
exact $R$-parity is required to keep the LSP stable if it is to be the 
DM.  Many WIMP candidates such as 
left-handed sneutrinos are disfavored \cite{Hall:1997ah} due to direct detection constraints. 

However, in order to accommodate the observed neutrino oscillations\footnote{See \cite{Bilenky:2004xm} for a review 
of the experimental evidence of neutrino oscillation and \cite{Cohen:2008qb}  
for a recent discussion of the theoretical formulation of neutrino oscillation.}
the SM should be supplemented with 
right handed neutrinos $\nu_R^i$.  The corresponding MSSM contains 
right-handed neutrino superfields that include both
the fermionic right-handed neutrinos and their scalar partners, the 
right-handed sneutrinos $\tilde{\nu}_R^i$. 
Further, any choice of scale for the Majorana mass for the RH neutrinos is technically natural, which leaves the 
Yukawa couplings in the neutrino sector a free parameter that can be varied in 
a wide range $10^{-13} \la y^i_\nu \la 1$.
If neutrino masses are Dirac (or nearly Dirac) 
then the Yukawa couplings are close to the lower end of this range, 
and the $\tilde{\nu}_R^i$ obtain the dominant contribution to their masses from 
the soft SUSY breaking mass terms in the MSSM;
two of the $\tilde{\nu}_R^i$ masses, resulting mostly from these mass terms,
could be the LSP and the NLSP.

Super-WIMP DM made up of $\tilde{\nu}_R^i$ 
was investigated recently in  
 \cite{Asaka:2007zz,Asaka:2006fs,McDonald:2006if} and found to be a viable option.\footnote{Decays of sneutrinos
 through R-parity violation was considered as a source of the PAMELA excess in \cite{Chen:2008dh}.} 
Indeed, $\tilde{\nu}_R^i$ are in some ways
a very promising candidate for DM in the galactic halo: 
they can have the right relic abundance \cite{Asaka:2007zz,Asaka:2006fs,McDonald:2006if}, 
and represents cold, neutral, colourless particles that do not 
interfere either with primordial nucleosynthesis or stellar evolution. 
Obviously, $\tilde{\nu}_R^i$ DM is consistent with limits on the self-interactions of DM and 
leaves no detectable signal in direct detection searches \cite{Taoso:2007qk,Bernabei:2008yi}
as their interactions are suppressed by $ y_\nu$. 

Despite these facts, one could question the phenomenological consequences of  light Dirac-like neutrino  masses
due to the required smallness of the yukawa coupling.  The SM Yukawa couplings already have a
hierarchy of $10^6$ between the smallest and largest Yukawas. 
This hierarchy, although puzzling, does not lead to 
problems with technical naturalness
as chiral symmetry is restored in the limit of 
vanishing Yukawa couplings. 
The Yukawa couplings for Dirac neutrinos are very small  $\mathcal{O}(10^{-13})$,
which leads to an even larger Yukawa coupling hierarchy
but are likewise technically natural. Further, Dirac neutrinos may not have lepton number violation 
which disfavors some forms of leptogenesis\footnote{Note that leptogenesis with dirac neutrinos can occur in some models, see \cite{Dick:1999je} for example.} as 
candidate theories for generating the observed matter-antimatter asymmetry of the Universe.
However, many alternatives for baryogenesis exist such as electroweak MSSM baryogenesis \cite{Carena:1996wj}, 
the Affleck-Dine mechanism \cite{Affleck:1984fy},  or an
effective modification of the EW phase transition \cite{Grojean:2004xa,Delaunay:2007wb,Grinstein:2008qi}.

The benefits of $\tilde{\nu}_R^i$ as DM arguably outweighs the theoretical 
costs for the reason mentioned in the introduction. 
If the PAMELA data is to be explained through the decay of a particle, then the decay width must be smaller than the width corresponding 
to the age of the universe $ \Gamma_U = \tau_U^{-1} \sim 10^{-42} \,  {\rm GeV}$. 
Such small widths are challenging to naturally produce for weak scale WIMPs.
On the other hand, in our scenario
the small decay width is linked to another very small 
number, namely $y_{\nu}^4 \sim (10^{-13})^4$, which is an interesting possibility.\footnote{This observation was made in
\cite{McDonald:2006if}, where it was pointed out that the lifetime
of the excited state of RH sneutrinos exceeds the
lifetime of the Universe.} 
One can immediately see that the decay  widths are such that $\Gamma_\nu \ll \Gamma_U$. 
These decays would still be 
occurring in the galactic halo leading to a primary source of positrons that 
could be the excess observed by PAMELA, and possibly explain the ATIC data as well.  

\subsection{Dirac $\tilde{\nu}_R^i$ as LSP and NLSP}

Following the initial stage of cosmological evolution, all superpartners of active 
SM species (charged under gauge groups) decay to $\tilde{\nu}_R^i$
on the time scales controled by $\Gamma \propto y_\nu^2\times({\rm weak~scale})$, see 
Refs. \cite{Asaka:2007zz,Asaka:2006fs} for details.
For our scenario, it is sufficient to assume that the population of the NLSP ($\tilde{\nu}_R^1$) 
is not parametrically small compared to the LSP ($\tilde{\nu}_R^2$) due to  
a very mild hierarchy of Yukawa couplings.

Consider the $R$-parity conserving MSSM supplemented with three 
Dirac $\tilde{\nu}_R^i$ with the superpotential
\begin{eqnarray}
W &=& y_{\nu}^{ij} \, \epsilon_{\alpha \, \beta} \,  \hat{H}_{u \, \alpha} \,  \hat{L}_{\beta}^j \, \hat{\nu}_R^i  
- \mu \, \epsilon_{\alpha \, \beta} \,  \hat{H}_{u \, \alpha} \,  \hat{H}_{d \, \beta} + \ldots
\end{eqnarray}
Here $\hat{H}_{u_,\alpha} = (\hat{H}_u^+,\hat{H}_u^0)$ and 
$\hat{H}_{d,\alpha} = (\hat{H}_d^0,\hat{H}_d^-)$ are the up and 
down type Higgs chiral superfields and $\hat{L}_{\beta} = (\hat{\nu}_L,\hat{\ell}^-_L)$ 
is the lepton chiral superfield and $\epsilon_{12} = 1$.  We have not written the 
terms in the superpotential containing the quark and charged lepton Yukawa couplings as 
they are not directly relevant to our discussion. 
The Yukawa interactions are
derived from the superpotential directly using
\begin{eqnarray}
\mathcal{L} = \frac12 \sum_{i,j} \frac{\partial^2 W}{\partial \phi_i \, 
\partial \phi_j} \psi_i \, \psi_j + h.c.
\end{eqnarray} 
Writing out the relevant component field terms in the resulting Lagrangian one has 
\begin{eqnarray}
\mathcal{L}_{yuk} &=&   y_\nu^{ij} \left(\tilde{\bar{\nu}}_R^i \, \nu^j \, 
\tilde{H}^0_u - \tilde{\bar{\nu}}^i_R \, \ell^j\, \tilde{H}_u^+ \right)  - \mu \left(\tilde{H}_u^+ \, \tilde{H}_d^- - \tilde{H}_u^0 \, \tilde{H}_d^0 \right) + h.c + \cdots
\end{eqnarray}
 
When the lowest lying supersymmetric particles are ($\tilde{\nu}_R^i,\tilde{\nu}_R^j$)
the decay of one of these species into another can proceed via 
$\tilde{\nu}_R^1 \rightarrow \tilde{\nu}_R^2 \, \bar \nu^k \, \nu^l $ or 
$\tilde{\nu}_R^1 \rightarrow \tilde{\nu}_R^2 \, \bar \ell^k \, \ell^l $. 
As we insist on exact $R$-parity, decays of $\tilde{\nu}_R^i$ to purely 
SM particles is forbidden, while all other supersymmetric final states are kinematically inaccessible.

\subsection{Leptonic $\tilde{\nu}_R^i$ decays}

\begin{figure}[hbtp]
\centerline{\scalebox{1}{\includegraphics{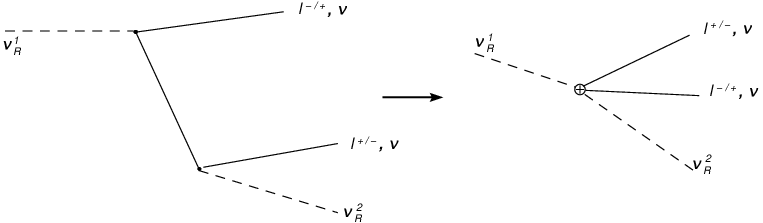}}}
\caption{The tree level decay process for $\tilde{\nu}_R^1
\rightarrow \tilde{\nu}_R^2 \, \ell^{\pm} \, \ell^{\mp}$, 
$ \tilde{\nu}_R^1 \rightarrow   \tilde{\nu}_R^2 \, \nu \, \nu$ 
and the local operator approximation.}
\label{sneutrinodecay}
\end{figure}
The leptonic decays  $\tilde{\nu}_R^1 \rightarrow \tilde{\nu}_R^2 \, \ell^k \, \ell^l $ 
proceed through Higgsino exchange\footnote{We disregard the issue of mixing between Higgsinos and gauginos as the mixing 
cannot introduce more than $O(1)$ corrections.} 
and the rate can be easily calculated as the function of 
$m_1$, $m_2$ and $\mu$, while masses of the SM leptons can be safely neglected. The amplitude is given by 
\begin{eqnarray}
i \mathcal{A} = i (y^{1j}_\nu \, y^{2i*}_\nu)  \, 
\bar L^j \, \frac{\xslash p  }{p^2 - \mu^2 } \, {L^i}.
\end{eqnarray}
Here $L$ is the left-handed lepton doublet, $1,2$ and $j,l$ are flavour indicies of RH sneutrinos and SM leptons respectively, 
and $p$ is the momentum carried by the Higgsino. 
Since all mass parameters are essentially free
apart from the $\mu > m_1>m_2$ constraint, after of quoting a general formula, 
we present the answer in several different limits. If the 
Higgsino mass $\mu$ is larger than the sneutrino mass scale, the Higgsino propagator 
can be contracted, as shown in Figure 1, and the decay is mediated at leading order in $p/\mu$ by the effective 
Lagrangian,
\be
{\cal L}_{\rm eff} = \fr{y^{1j}_\nu \, y^{2i*}_\nu}{\mu^2}\, (\tilde{\nu}_R^{2*}\partial_\mu \tilde{\nu}_R^{1})\,
\bar L^j \gamma_\mu L^i,
\ee
which is simply a product of sneutrino and left-handed lepton currents. Introducing 
effective couplings $y_{1}^2 = \sum_j |y^{1j}_\nu|^2, \, ~y_{2}^2 = \sum_j |y^{2j}_\nu|^2 $
and the energy release parameter $\Delta = m_1-m_2$ one can calculate the general decay width in a straightforward
manner. One finds the result for the decay width given by
\begin{eqnarray}
\frac{\Gamma_{l^+ \, l^-}}{y_1^2\, y_2^2}  &=& -\frac{\Delta  \left(24  \, m_1^7-84 \Delta  m_1^6 + 104 \Delta ^2 m_1^5-50 \Delta^3 m_1^4-8 \Delta ^4 m_1^3+20 \Delta^5 m_1^2-8 \Delta ^6 m_1+\Delta^7\right)}{6144 \, m_1^3 \, \pi ^3 \, \mu ^4}, \nn \\
&\,&  - \frac{12 \, m_1^4 \, \log \left[\frac{(m_1- \Delta)^2}{m_1^2} \right] \, \left(m_1 - \Delta \right)^4}{6144 \, m_1^3 \, \pi ^3 \, \mu ^4}.
\end{eqnarray}

This general result obscures the physics somewhat, however, recall that in our scenario the sneutrinos are 
the LSP and the NLSP so necessarily the Higgsino is a larger mass scale. Let $\mu = n \, m_1$ where $n$ is an 
order one number ($n>1$) for this reason. Further, for this decay to occur $m_2 < m_1$ so let $m_2 = x \, m_1$ where 
$x < 1$, then expanding in small $x$ one obtains
\begin{eqnarray}
\Gamma_{l^+ \, l^-} &=& \frac{m_1 \, y_1^2\, y_2^2}{6144 \, n^4 \, \pi^3}\left(1 - 8\, x^2 - 24 \, x^4 \, \log \left(x \right) + 8 x^6 + \mathcal{O}(x^7)\right). 
\end{eqnarray}
Thus we see that a good order of magnitude approximation for this decay width is $\Gamma_{l^+ \, l^-}  \simeq 10^{-6}  \, y_1^2 \, y_2^2 \, m_1$ for the whole range of parameter space. 

We also note that this decay width has a number of phenomenologically interesting limits. For example,
considering $\Delta$ to be somewhat small on the scale of the sneutrino masses, we
derive the total decay rate into charged leptons,
\be
\Gamma_{l^+ \, l^-}^a = \fr{y_1^2\, y_2^2\, \Delta^5}{480\,\pi^3\,\mu^4}~~~{\rm for}~~ \Delta \ll m_{1,2} \ll \mu.
\label{first_ap}
\ee
This formula can be obtained directly from the neutron beta decay rate upon setting
$m_e,g_A\to 0$, and identifying $G_F\cos\theta_c/ \sqrt{2} \to 1/ (4\, \mu^2)$. 
In another kinematic regime, when $\mu$ is very close to $m_1$ and the energy release remains 
small, the rate is given by
\be
\Gamma_{l^+ \, l^-}^b = \fr{y_1^2\, y_2^2\, \Delta^3}{192 \,\pi^3\,m_1^2}~~~{\rm for}~~ \mu-m_1\ll \Delta \ll m_{1,2}.
\label{second_ap}
\ee
Finally, if the scale of $m_2$ is small and can be neglected, while $\mu$ remains close to $m_1$
we have a third limit,
\be
\Gamma_{l^+ \, l^-}^c = \fr{y_1^2\, y_2^2\, m_1}{1024\,\pi^3}~~~{\rm for}~~ \mu-m_1,~m_2 \ll m_{1}.
\label{third_ap}
\ee
We do not require that any of the hierarchies between mass parameters $m_1,~m_2,~\mu$ used to determine the widths $\Gamma^{a,b,c}_{l^+ \, l^-}$  is strickly realized in our scenario. We have determined $\Gamma^{a,b,c}_{l^+ \, l^-}$
using these limits on  $m_1,~m_2,~\mu$ to determine some phenomenologically interesting limits of the leptonic decay width. In what follows we will use the approximation $\Gamma_{l^+ \, l^-}  \simeq 10^{-6} y_1^2 \, y_2^2 m_1$ which is a good approximation due to the physics of our scenario, (ie a Higgsino heavier that the LSP and NLSP sneutrinos and $m_2 < m_1$).

\subsection{Hadronic $\tilde{\nu}_R^i$ decays
as a source of $\bar{p}$ flux }

Depending on kinematics, besides the pure leptonic decay channels, there could be other
decay channels potentially leading to energetic anti-quarks and eventually to anti-protons.
Quark-antiquark pairs cannot be produced in the sneutrino decay via the intermediate 
gauge and/or Higgs bosons. 
The relevant part of the scalar potential that could lead to such decays is
\be
V = 2 \, \mu \, y_\nu^{ij} \,  \tilde{\nu}_R^i (\tilde{\nu}^j_L \, H_d^0 -  \tilde{\ell}^j_L \, H_d^+) + 
\sum_k y_\nu^{ik} \, (y_\nu^{jk})^\star H_u^0 \, (H_u^0)^\star \tilde{\nu}_R^i  \tilde{\nu}_R^j + h.c +...
\label{scpot}
\ee
These interactions induce three- and four-body decays of RH neutrinos: 
$\tilde{\nu}_R^1 \rightarrow \tilde{\nu}_R^2 \, h(H)$, 
$\tilde{\nu}_R^1 \rightarrow \tilde{\nu}_R^2 \, W^+W^-(H^+H^-)$, 
$\tilde{\nu}_R^1 \rightarrow \tilde{\nu}_R^2 \, ZZ(AA)$,  decays.
To simplify our calculations, we shall assume that $\tan\beta = 
\langle H_u^0 \rangle/ \langle H_d^0 \rangle$
is somewhat large, and the $A,H,H^{\pm}$ Higgs bosons are heavy,
so that only the decay to the lightest Higgs boson $h$ and pairs of gauge bosons 
are kinematically possible. 

We concentrate on the three-body decay $\tilde{\nu}_R^i \rightarrow \tilde{\nu}_R^j \, h$,
which is driven by the last term in Eqn. (\ref{scpot}) when one of the Higgs is given a vev. The resulting decay width is
\be
\Gamma_h = |y_{12}^2|^2 \, \fr{v^2}{8\pi m_1^2}\times v_h \left(\Delta +\fr{m_h^2-\Delta^2}{2 \, m_1}\right).
\label{twobody}
\ee
In this formula, $v_h$ is the velocity of the outgoing Higgs boson, $v=246$ GeV is the SM Higgs vev, 
and $y_{12}^2$ is the sneutrino-flavor changing combination of the Yukawa couplings:
\be
y_{12}^2 \equiv \sum_j y_\nu^{1j} \, (y_\nu^{2j})^\star.
\ee


If $m_1 \sim v$, $v_h \sim 1$ and $y_{12}^2\sim y_1y_2$, 
the two-body decay rate would dominate over the leptonic rate by three orders of magnitude
simply because of the phase space suppression of Eqs.(\ref{first_ap})-(\ref{third_ap}) relative to Eqn.(\ref{twobody})
is quite significant. The Higgs boson produced in the decay fragments further to gauge bosons and 
heavy quarks and leptons. For SUSY models its typical 
mass precludes it from decaying directly to $W^+W^-$ or $ZZ$, and $b\bar b$ decays are expected to dominate. 
Decays of this form is often encountered in neutralino annihilation, and the corresponding 
yield of antiprotons from a $b$-quark injection has been evaluated
\cite{Bottino:1994xs,Bottino:2005xy}. The yield of antiprotons in the hadronization and 
subsequent decays of $b\bar b$-pairs is quite significant, exceeding 10\% per annihilation/decay 
event with a typical electroweak scale energy injection. Assuming an $\mathcal{O}(0.1-1)$ yield of 
antiprotons in the $\tilde{\nu}_R^1 \rightarrow \tilde{\nu}_R^2 \, h$ decay, the ratio of 
Eqn.(\ref{twobody}) to a typical three-body rate gives the relative strengths of antiprotons
and positrons at the injection:
\be
\fr{\Phi_{\bar p}}{\Phi_{e^+}} \sim 10^3 \times \fr{|y_{12}^2|^2 }{y_1^2\, y_2^2}\times v_h.
\label{ratio}
\ee 
A recent theoretical study of the antiproton fraction \cite{Donato:2008jk} 
limits the "boost factor" for the annihilation 
of dark matter to be less than 6(40) for the annihilation of 0.1(1) TeV particles. 
If we tune the model to fit PAMELA flux for positrons, this would translate 
into a constraint on Eqn.(\ref{ratio}) to be smaller than 0.1-1. Consequently,
one would have to require a suppression of $v_h |y_{12}^2|^2/(y_1^2\, y_2^2) $ 
down to the level of $10^{-4}-10^{-3}$. Such a suppression may come from the smallness of 
the flavor-changing coupling $y_{12}^2$,
or may originate kinematically, from small scale splitting, $\Delta < m_h$.
In both cases one would be able to fit PAMELA results without overproducing 
antiprotons.

Note that decays that are loop suppressed will generally also have the difference in flavour decays to quarks and thus protons and anti-protons compared to leptonic decays. This is due to the exact R parity of our scenario and the fact that only the LSP $\tilde{\nu}_R^2$ is kinematically accessible for the NLSP $\tilde{\nu}_R^1$ to decay too. Loop decays with the same flavour structure do exist where the produced leptons form a loop that produces a $Z/\gamma$ that subsequently decays to hadrons. The most problematic of these decays are suppressed by 
\be
\fr{\Phi_{\bar p}}{\Phi_{e^+}} \sim 10^3 \times \fr{g^2 \, g_V^2}{16 \pi^2 ( 4 \cos^2(\theta_W))}\times BR(Z \rightarrow {\rm hadrons})
\ee 
and the suppression is smaller than the required 0.1-1.

\subsection{Numerology}

Our main result, Eqs.(\ref{first_ap})-(\ref{third_ap}) can give an excellent fit to the PAMELA data in agreement 
with the requirement of Eqn.(\ref{decay}). Adopting $\Gamma_{l^+ \, l^-} = 10^{-6} \times y_1^2\, y_2^2 \times m_1$ 
as an approximation for different kinematic regimes of the three-body decay rate, 
we conclude that the Yukawa couplings must satisfy the following relation:
\begin{eqnarray}
 (y_1 \, y_2)^2  \sim  10^{-52} ~~\Longrightarrow~~ (y_1 \, y_2)^{1/2} \sim 1\times 10^{-13}.
 \label{whatyouget}
\end{eqnarray}

Are these values for the Yukawa couplings in agreement with Dirac neutrino Yukawa couplings suggested
by the measurements of neutrino oscillations? The flavor 
basis in the RH sneutrino sector and the active neutrino sector in general do not coincide
and a one-to-one connection between neutrino phenomenology and Eqn.(\ref{whatyouget}) is not possible.
The flavor-changing combination of Yukawa couplings 
in Eqn.(\ref{twobody}), remains uncertain and is not necessarily related to the measured 
mixing angles.
If further assumptions are made about the structure of the PMNS and the $V_R$ mass matricies
in a particular constrained MSSM then the the flavour structure we have determined can be related to the 
PMNS mass matrix, but this is beyond the scope of this work. 
No immediate connection between our determined flavour structure and the obervable masses and mixing angles
in the neutrino sector is possible without further experimental input, but we can check the consistency of Eqn.(\ref{whatyouget}) with general expectations 
for the Yukawa couplings in the Dirac neutrino sector. 
Recall that Dirac neutrinos obtain their masses via 
$ m_\nu = y_\nu \langle H_u^0 \rangle = y_\nu \, v \, \sin \beta/\sqrt{2} \simeq y_\nu \times 175$ GeV,
as we assume $\tan\beta$ to be large.  
To estimate the possible size of the Yukawa couplings we use the experimental 
determinations of the neutrino mass splitting. 
The difference in
masses squared of the neutrinos have been measured by the 
KamLAND \cite{Araki:2004mb} and K2K \cite{Aliu:2004sq} collaborations to be 
\begin{eqnarray}
[\delta m_\nu^2]_{\rm atm} \simeq 2.8 \times 10^{-3} \, {\rm eV}^2, 
\quad [\delta m_\nu^2]_{\rm solar} \simeq 7.9 \times 10^{-5} \, {\rm eV}^2 
\end{eqnarray}
and the absolute bound on the sum of neutrino masses from WMAP is given to be 
\begin{eqnarray}
\sum m_\nu^i = 0.67 \, {\rm eV} \, {\rm at} \, 95 \, \% \,  {\rm CL}.
\end{eqnarray}
which coincides with the limit on the heaviest neutrino mass 
$ m_\nu < 0.7 {\rm eV}$.

In the case of the standard hierarchy of neutrino masses  
the heaviest neutrino is given by $\sim \sqrt{(\delta \, m_\nu^2)_{\rm atm}}$ 
and one finds that the largest Yukawa coupling is  
\begin{eqnarray}
 y_\nu  \simeq 3.0 \, \times 10^{-13} \, 
\left(\frac{m_\nu^2}{2.8 \times 10^{-3} \, {\rm eV}^2} \right)^{1/2},
\end{eqnarray} 
which is perfectly consistent with the requirement on $(y_1 \, y_2)^{1/2}$ in our scenario. 
Alternatively, for the small neutrino mass splitting scenarios one 
could have Yukawa couplings close to the value suggested by
possible given cosmological bounds, $ y_\nu  \simeq  10^{-12}$. In either case
we see that the Yukawa couplings can realize our super-WIMP decay scenario that 
produces the PAMELA,  and possibly ATIC, signals for electroweak scale DM masses.

\section{Discussion and Conclusions}

We have presented an observation that the decay rates of super-WIMPs with the rate 
about ten orders of magnitude slower than the Hubble rate may originate from 
the transition of one super-WIMP state into another in the {\em second} order
in the super-WIMP coupling so that $\Gamma \sim y_{SW}^4\times({\rm weak~scale})$. 
The resulting decay rates are just becoming possible to probe through 
indirect detection of dark matter via its decay products, such as electrons, positrons and antiprotons.

Perhaps the most interesting aspect of our model is that 
super-WIMP physics is extremely sensitive to the superpartner mass 
spectrum due to the existence of a very small coupling $y_{SW}$. If, for example, one has a mass
spectrum such that
$m_{\tilde \nu_{R}^{2}} < m_{\rm Higgsino} < m_{\tilde \nu_{R}^{1}}$,
the sequential decays $\tilde \nu_{R}^{1}\to {\rm Higgsino}\to \tilde \nu_{R}^{2}$ 
will happen with $\Gamma \propto y_\nu^2 \times ({\rm weak~scale})\propto 10~{\rm Hz}$, which is almost 
instantaneous on the scale of Eqn.(\ref{decay}). A simple 
modification of the spectrum to $m_{\tilde \nu_{R}^{1}} < m_{\tilde \nu_{R}^{2}} < m_{\rm Higgsino} $
was shown to leads to an enormous delay in the $\tilde \nu_{R}^{1}\to \tilde \nu_{R}^{2}$ decay. 
This delay results in an overall drop in the decay rate by 25 orders of magntiude, 
and makes the heavier component of the RH sneutrinos only very weakly unstable. 
The numbers, given a huge disparities of different scales involved, work remarkably well, 
producing $\Gamma_{e^+} \sim m_\nu^4/({\rm weak~scale})^3$, which is in agreement with a
putative explanation of PAMELA (and ATIC) signals by the decaying super-WIMPs. Moreover, for a moderate
splitting between two RH sneutrino components and/or an accidental suppression of 
the different flavor transition, the hadronic 
decays of $\tilde \nu_{R}^{1}$ will be suppressed, which will in turn suppress the 
antiproton flux created by super-WIMPs. A kinematic suppression of the antiproton 
flux would imply a rather small energy splitting, $ \Delta < m_h\sim O(100)$ GeV
between two super-WIMP states. Alternatively, if both PAMELA and ATIC are to be explained 
in the same way, one would have to choose $\Delta$ close to a TeV and 
assume a hierarchy  $|y_{12}^2|^2\ll y_1^2y_2^2$ in order to suppress the antiproton flux. 

One feature of our scenario may look somewhat unusual. We do not assume a complete degeneracy 
of RH sneutrino masses, and in fact requite them being split by the energy 
intervals comparable to their masses. This is not possible in the context of universal
sfermion masses often employed in the SUSY literature, as any additional splitting induced 
by $y_\nu$ is minuscule. Therefore, one would have to imagine some 
additional theoretical mechanism to create such a splitting. 

Although the example given in this note is quite natural, it is hardly unique. For example, 
one could envisage sequential $R$-parity preserving decays of other super-WIMPs 
featured in the SUSY literature, such as gravitinos, axinos, modul(inos) etc. 

The decaying super-WIMP scenario as a tentative explanation of PAMELA and ATIC
differs from the boosted WIMP annihilation in many ways. The most notable 
distinction is, of course, the early cosmology. The enhanced annihilation 
may lead to the nuclear-chemical consequences such as the overproduction of $^6$Li \cite{Jedamzik:2004ip},  
and to extra ionization and a diffuse gamma background 
\cite{Kamionkowski:2008gj} produced by the annihilating WIMPs. In contrast, the decaying DM 
scenario is immune to these potential problems. On the other hand, both annihilation and decay scenarios
are subject to gamma ray and synchrotron emisson constraints (for a recent paper see  \cite{Nardi:2008ix}) coming from the 
central region of the galaxy where the density of dark matter is significantly 
enhanced. 

\subsection*{Acknowledgements}

We would like to thank Adam Ritz, Brian Batell and Vivek Sharma for useful discussions. 
The work of MP was supported in part by NSERC, Canada. Research at the Perimeter Institute
is also supported in part by the Government of Canada through NSERC and by the Province
of Ontario through MEDT. 

\bibliography{DM}
\end{document}